\documentclass[]{aa}  
\usepackage{subfigure}
\usepackage{graphicx}
\usepackage{xcolor}
\usepackage{txfonts}
\usepackage{natbib,twoopt}
\usepackage[breaklinks=true]{hyperref} 
\bibpunct{(}{)}{;}{a}{}{,}             
\makeatletter
  \newcommandtwoopt{\citeads}[3][][]{\href{http://adsabs.harvard.edu/abs/#3}%
    {\def\hyper@linkstart##1##2{}%
     \let\hyper@linkend\@empty\citealp[#1][#2]{#3}}}
  \newcommandtwoopt{\citepads}[3][][]{\href{http://adsabs.harvard.edu/abs/#3}%
    {\def\hyper@linkstart##1##2{}%
     \let\hyper@linkend\@empty\citep[#1][#2]{#3}}}
  \newcommandtwoopt{\citetads}[3][][]{\href{http://adsabs.harvard.edu/abs/#3}%
    {\def\hyper@linkstart##1##2{}%
     \let\hyper@linkend\@empty\citet[#1][#2]{#3}}}
  \newcommandtwoopt{\citeyearads}[3][][]%
    {\href{http://adsabs.harvard.edu/abs/#3}
    {\def\hyper@linkstart##1##2{}%
     \let\hyper@linkend\@empty\citeyear[#1][#2]{#3}}}
\makeatother

\begin{document} 

\defcitealias{pp24}{P24} 

   \title{The hot gas mass fraction in halos. \\From Milky Way-like groups to massive clusters}


   \author{P. Popesso\inst{1,}\inst{2}\thanks{paola.popesso@eso.org}
   \and A. Biviano\inst{3,}\inst{4}
   \and I. Marini\inst{1}
    \and K. Dolag\inst{5,}\inst{6,}\inst{2}
    \and S. Vladutescu-Zopp\inst{5}
    \and B. Csizi\inst{7}
    \and V. Biffi\inst{8}
    \and G. Lamer\inst{9}
    \and A. Robothan\inst{10}
    \and M. Bravo\inst{11}
    \and L. Lovisari\inst{12,}\inst{13}
    \and S. Ettori\inst{14}
    \and M. Angelinelli\inst{14}
    \and S. Driver\inst{9}
    \and V. Toptun\inst{1}
    \and A. Dev\inst{10}
    \and D. Mazengo\inst{1}
    \and A. Merloni\inst{15}
    \and J. Comparat\inst{15}
    \and G. Ponti\inst{15}
    \and T. Mroczkowski\inst{1}
    \and E. Bulbul\inst{15}
    \and S. Grandis\inst{7}
    \and E. Bahar\inst{15}
          }

   \institute{European Southern Observatory, Karl Schwarzschildstrasse 2, 85748, Garching bei M\"unchen, Germany \email{paola.popesso@eso.org}
         \and
            Excellence Cluster ORIGINS, Boltzmannstr. 2, D-85748 Garching bei M\"unchen, Germany
            \and
                INAF – Osservatorio Astronomico di Trieste, Via Tiepolo 11, 34143 Trieste, Italy
        \and
            IFPU – Institute for Fundamental Physics of the Universe, Via Beirut 2, I-34014 Trieste, Italy
            \and
             Universitäts-Sternwarte, Fakultät für Physik, Ludwig-Maximilians-Universität München, Scheinerstr.1, 81679 München, Germany
        \and 
            Max-Planck-Institut für Astrophysik, Karl-Schwarzschildstr. 1, 85741 Garching bei M\"unchen, Germany
                    \and
        Universität Innsbruck, Institut für Astro- und Teilchenphysik, Technikerstr. 25/8, 6020 Innsbruck, Austria
               \and
             INAF – Osservatorio Astronomico di Trieste, Via Tiepolo 11, 34143 Trieste, Italy
        \and
            Leibniz-Institut für Astrophysik Potsdam (AIP), An der Sternwarte 16, 14482 Potsdam, Germany   
        \and
            International Centre for Radio Astronomy Research, University of Western Australia, M468, 35 Stirling Highway, Perth, WA 6009, Australia
            \and 
            McMaster University, 1280 Main Street West, Hamilton, Ontario, L8S 4L8, Canada
        \and
            INAF– Osservatorio Astronomico di Brera, Via E. Bianchi 46, 23807 Merate (LC), Italy
            \and
            Center for Astrophysics $|$ Harvard $\&$ Smithsonian, 60 Garden Street, Cambridge, MA 02138, USA
        \and
            INAF– Osservatorio Astronomico di Bologna, Via Gobetti 93/3, 40129 Bologna, Italy
            \and
            Max-Planck-Institut für Extraterrestrische Physik (MPE), Giessenbachstr. 1, D-85748 Garching bei München, Germany
             }

   \date{Received September 15, 1996; accepted March 16, 1997}

  \abstract
   {}
   {By using eROSITA data in the eFEDS area, we provide a measure of the $f_{gas}-M_{halo}$ relation over the largest halo mass range, from Milky Way sized halos to massive clusters, and to the largest radii ($R_{200}$) ever probed so far in local systems at $z< 0.2$.}
   {To cope with incompleteness and selection biases of the X-ray selection, we apply the stacking technique in eROSITA data of a highly complete and tested sample of optically selected groups. The method has been extensively tested on mock observations.}
   {In massive clusters, the hot gas alone provides a baryon budget within $R_{200}$ consistent with $\Omega_{b}/\Omega_{m}$, while at the group mass scale, it accounts only for 20-40\% of the cosmic value. The $f_{gas}-M_{halo}$ relation is well fitted by a power law with a consistent shape (within 1$\sigma$) at $R_{500}$ and $R_{200}$ and a normalization varying nearly by a factor of 2. Such a relation is consistent with other works in the literature that consider X-ray survey data at the same depth as eFEDS, but it provides a lower average $f_{gas}$ in the group regime in comparison to works based on X-ray bright group samples. The comparison of the observed relation with the predictions of several hydrodynamical simulations (BAHAMAS, FLAMINGO, SIMBA, Illustris, IllustrisTNG, MillenniumTNG and {\it Magneticum}) shows that all state-of-the-art simulations but {\it Magneticum} over-predict the gas fraction, with the largest discrepancy (up to a factor 3) in the  $10^{13.5}$ $M_\odot$ to $10^{14.5}$ $M_\odot$ halo mass range.}
   {We emphasize the need for mechanisms that can effectively expel gas to larger radii in galaxy groups without excessively quenching star formation in their member galaxies. Current hydrodynamical simulations face a significant challenge in balancing their subgrid physics: none can sufficiently evacuate gas from the halo virial region without negatively impacting the properties of the resident galaxy population.}


   \maketitle
%

\section{Introduction}
In a purely gravitational framework, the temperature, density, and mass distribution of the hot gas in galaxy groups and clusters are expected to follow the halo’s potential well and the dark matter distribution in a self-similar regime, as predicted by large-scale structure evolution models \citep[see][]{Peebles1980}. However, observations consistently show significant deviations from self-similar predictions in various scaling relations \citep{Popesso2005, rykoff2008, Vikhlinin09, pratt09, Mantz2015, Planck2016, bulbul19, Bahar22, Zhang24b}. Specifically, several studies indicate that the baryon fraction within the central region ($<R_{500}$\footnote{$R_{\Delta}$ represents the radius of a sphere centered on the group, with a mean density equal to $\Delta$ times the critical density of the Universe at the group’s redshift.}) of massive halos increases with halo mass \citep{Sun2009, Ettori15, Lovisari2015, Eckert16, Nugent20}. The baryon content in galaxy groups is only about half of the expected value based on self-similar models, while for clusters it is much closer to the predicted value \citep[see, e.g., the review by][]{Eckert21}. This discrepancy is thought to be driven by non-gravitational processes that alter the thermodynamic properties of the hot gas and the overall baryon content in groups and clusters. Among these processes, feedback from the central supermassive black hole (hereafter BH) in the brightest cluster galaxy is a strong candidate due to the substantial energy involved \citep{Sijacki07, Puchwein08, Fabjan10, McCarthy10, LeBrun2014, biffi18, VallesPerez20, Galarraga-Espinosa21, Eckert21, Oppenheimer21}. In high-mass clusters, BH feedback primarily impacts the central core, leading to elevated entropy levels \citep{LeBrun2014}. In contrast, in lower-mass groups, BH feedback may affect the gas out to 0.5-1 Mpc, potentially influencing the entire group volume \citep[e.g., see][for a comprehensive review]{Oppenheimer21}.

Cosmological simulations show that different implementations of BH feedback result in varying predictions, particularly at the group mass scale. In single-mode feedback models, a fraction of the BH energy is deposited as a thermal boost into nearby cells \citep[e.g.,][]{Schaye15, McCarthy17, Schaye23}. More sophisticated dual-mode implementations distinguish between thermal outflows or high-accretion (quasar) modes and kinetic energy transfer or bubble inflation at low accretion rates (radio mode), incorporating halo mass dependencies in models such as {\it IllustrisTNG}\citep{pillepich19}, {\it MillenniumTNG} \citep{Pakmor22}, and {\it Magneticum} \citep{Dolag16}. Constraining these theoretical predictions requires precise observational measurements of the hot gas content in galaxy groups and clusters to place upper limits on the energy injected via BH feedback.

Current observational studies of the gas mass fraction ($f_{gas}$) as a function of halo mass ($M_{halo}$) primarily focus on clusters. At lower masses, only a limited number of heterogeneously selected X-ray groups are available, sampled at varying depths and resolutions \citep{Ponman1996, Mulchaey2000, Osmond2004, Sun2009, Lovisari2015, 2021Univ....7..254L}. Pointed observations with ROSAT, XMM-Newton, and Chandra have targeted only the brightest X-ray groups. If BH feedback can expel part of the intragroup medium, it would reduce gas density and potentially bias flux-limited X-ray surveys, leading to an overestimation of $f_{gas}$ at the group scale. Therefore, an unbiased selection method is needed to accurately measure the hot gas content in galaxy groups. Indeed, recent analyses of eROSITA Science Verification data over the eROSITA Final Equatorial Depth Survey (eFEDS) area \citep{Brunner2022} suggest that eROSITA’s X-ray selection captures only a subset of galaxy groups, with many remaining undetected due to their lower X-ray surface brightness at a given halo mass \citep[][P24 hereafter]{pp24}.

An optical selection of galaxy groups offers a way to circumvent BH feedback biases since it is independent of a system's hot gas content. Large spectroscopic surveys like SDSS \citep{blanton2017} and GAMA \citep{Driver2022} provide comprehensive catalogs of optically selected groups and clusters, down to halo masses of $10^{12}$ $M_{\odot}$ at low to intermediate redshifts \citep[e.g.,][]{Tempel2017, Robotham2011, 2007ApJ...671..153Y}. Combining such optically selected samples with large X-ray surveys has proven effective in characterizing the average X-ray properties of the halo population down to group scales \citep{Anderson2015, rozo2009, rykoff2008, Crossett2022, Giles2022}. Stacking X-ray data at the positions of optically detected groups enables the measurement of average X-ray scaling relations in the low-mass regime. In this paper, we perform a stacking analysis on eROSITA data of the GAMA galaxy group sample. The methodology and selection effects arising from the optical selection and the halo mass proxy used as a prior for the stacking have been thoroughly tested and evaluated in previous studies \citep[][]{Popesso2024b,Marinia}. This study offers the most precise estimate of the gas mass fraction to date, covering the broadest halo mass range and the largest radius ever examined, while minimizing biases related to BH feedback, after extensive testing on mock dataset.

The paper is structured as follows. In Sect.~\ref{dataset} we describe optical and X-ray datasets used in the analysis. In Sect.~\ref{fgas} we describe how the gas mass is derived from the X-ray surface brightness distribution obtained through stacking in \citetalias{pp24}. Sect.~\ref{res} provides our results, while in Sect.~\ref{conc} we draw our conclusions. Throughout the paper we assume a Flat $\Lambda$CDM cosmology with $H_0 = 67.74$ km~s$^{-1}$~Mpc$^{-1}$, $\Omega{_m}(z = 0) = 0.3089$ \citep{Planck2016}.

\section{The dataset}
\label{dataset}
In this section, we describe the data available over the 60 deg$^2$ of overlapping area between the eFEDS and the GAMA survey. 

\subsection{eROSITA eFEDS data and group sample}\label{erosita}

For this study, we used the public Early Data Release (EDR) eROSITA event file of the eFEDS field \citep{Brunner2022}. The field was observed with an unvignetted exposure of approximately 2.5 ks, slightly higher than the anticipated exposure for the future eRASS upon completion, which is about 1.6 ks unvignetted. The dataset contains roughly 11 million events (X-ray photons) detected by eROSITA across the 140 deg$^2$ area of the eFEDS Performance Verification survey. Each photon is assigned an exposure time based on the vignetting-corrected exposure map. Photons in proximity to detected sources from the source catalog are flagged. These sources are classified as point-like or extended according to their X-ray morphology \citep{Brunner2022} and further categorized (e.g., galactic, active galactic nuclei, individual galaxies at redshift $z < 0.05$, galaxy groups, and clusters) using multi-wavelength information \citep{Salvato2022, Vulic2022, LiuTeng2022, LiuAng2022, Bulbul2022}. 

The X-ray-selected groups and clusters in the eFEDS area are provided in the catalog of \citet{LiuAng2022}. This comprises more than 500 extended objects up to z$\sim$1. According to \citet{LiuAng2022} the catalog reaches a completeness of 40\% down to a flux limit of $1.5 \times 10^{-14}$ erg~s$^{-1}$~cm$^{-2}$. Each source is assigned a redshift according to the analysis reported in \cite{Klein2022}, an estimate of the X-ray luminosity within different apertures (300 and 500 kpc and within $R_{500}$), and the surface brightness distribution within several radii. An estimate of $M_{500}$\footnote{$G \, M_{\Delta}=\Delta/2 \, H_z^2 \, R_{\Delta}^3$, where $G$ is the gravitational constant and $H_z$ the Hubble constant at the group redshift $z$.} is provided on the basis of the $L_X-M_{500}$ scaling relation of \citet{Lovisari2015}.

Despite the smaller volume covered by eFEDS, its much greater depth and stable background make it preferable to the shallower observations from eRASS:1. Therefore, we use the results of \citet{Popesso2024c} on the average X-ray surface brightness profiles of galaxy groups to derive gas mass profiles from Milky Way-sized halos to massive clusters. These profiles, along with the stacking technique they are based on, have been rigorously validated by \citet{Popesso2024b} using mock optically selected catalogs and simulated eROSITA observations. This testing was done to assess the contamination and completeness of the input sample and to verify the reliability of the stacking analysis.


\subsection{The GAMA optically selected group and cluster sample}\label{gama}
The optically selected GAMA group sample \citep{Robotham2011} comprises about 7500 galaxy groups and pairs identified in the spectroscopic sample of the GAMA spectroscopic survey \citep{Driver2022}, over the region of interest (G09 in the GAMA survey). This reaches a completeness of $\sim 95\%$ down to the magnitude limit of $r=19.8$. The Friends-of-Friends algorithm described in \cite[][hereafter R11]{Robotham2011} identifies the galaxy groups and pairs. 

Once galaxy membership is identified, the mean group coordinates and redshift are iteratively estimated for each system. The total mass of the systems ($M_{\mathrm{fof}}$) is then estimated from the group’s velocity dispersion ($\sigma_v$) within a variable radius \citep[see][for more details]{Robotham2011}. However, \citet{Marini24b}, using a mock synthetic catalog based on the same selection algorithm of \citet{Robotham2011}, show that the halo mass proxy based on velocity dispersion is not a reliable measure for groups with a low number of galaxy members. To address this, a richness cut is required to ensure a minimum number of galaxies to accurately estimate the dispersion, which introduces further selection effects during stacking. 

\citet{Marini24b} indicate that the total optical luminosity of the groups is the best mass proxy. The algorithm effectively retrieves group membership, ensuring that all bright members contributing most to the group’s total luminosity are captured. This accuracy is consistent regardless of group richness, including the case of pairs. Therefore, the halo mass proxy based on total luminosity not only shows the best agreement with the true/input halo mass but also allows for the creation of a clean group sample without additional selection effects due to richness cuts.

Consequently, we use the galaxy membership provided in \citet{Robotham2011}'s catalog to estimate the total group optical luminosities in the r-band following the approach of \cite{Popesso2005}, and the group masses $M_{200}$ and virial radii $R_{200}$ from the scaling relations with the optical luminosity provided in the same paper. Additionally, we derive estimates for $M_{500}$ and $R_{500}$ using the NFW mass distribution model of \citet{NFW1997} and the concentration-mass relation of \citet{DM14}.

\begin{figure*}
\begin{center}

\includegraphics[width=0.85\textwidth]{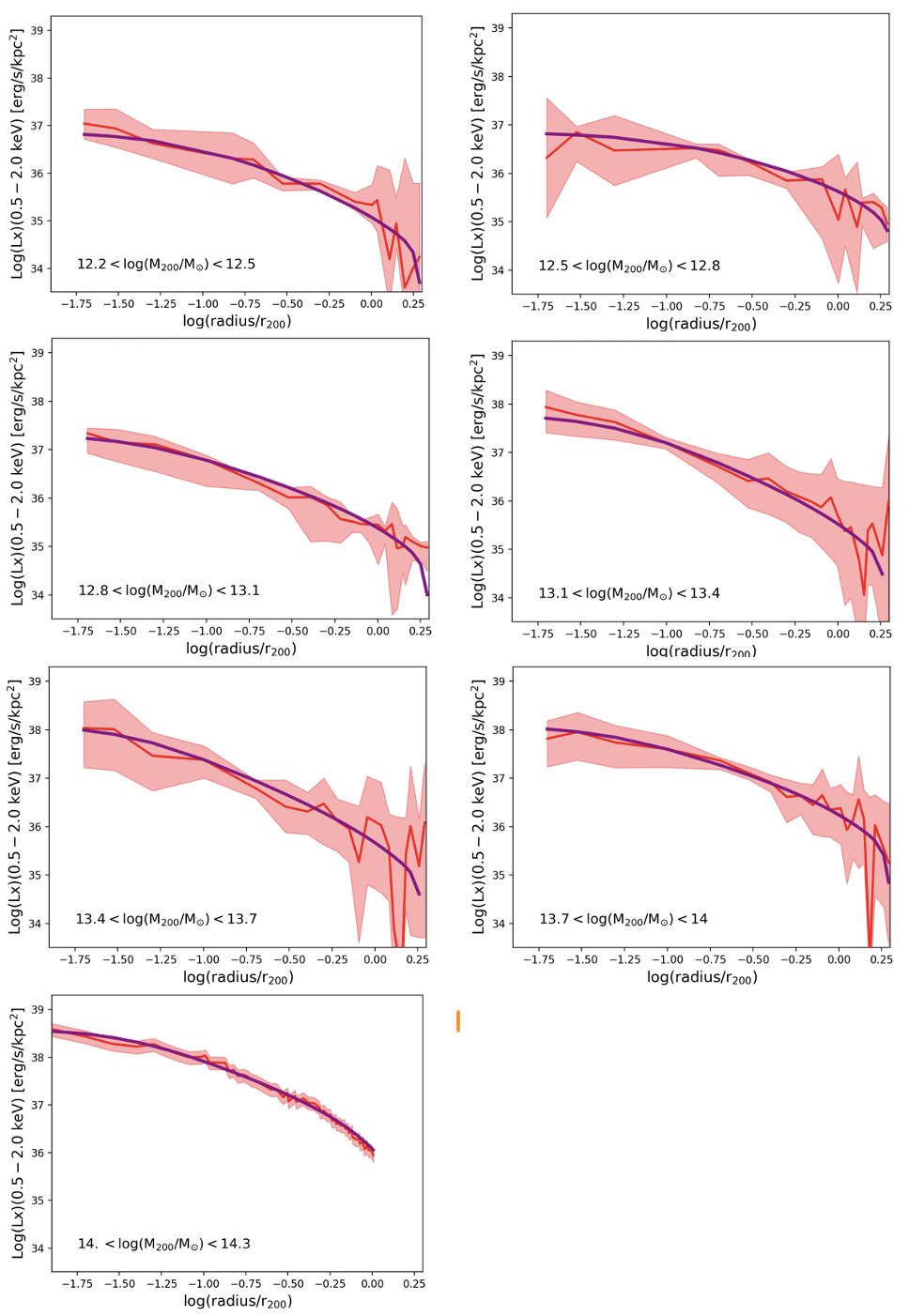}
\caption{Each panel in the upper two rows shows the X-ray surface brightness profile as estimated in \citet{Popesso2024c} The red solid line represents the stacked profile, while the pink shaded region indicates the uncertainty derived from bootstrapping. The panel in the last row displays the average X-ray surface brightness profile (red solid line) of the eFEDS detections with optical counterparts in the GAMA galaxy sample, corresponding to the specific halo mass bin, while the pink shaded region indicates the dispersion. The solid purple line in all panels shows the best-fit projected X-ray emissivity profile, represented by $n_e(r)^2\Lambda(kT, Z)$.}
\label{a1}
\end{center}
\end{figure*}

\begin{table*}
\caption{Bets fit electron density model parameters of the \citet{Vikhlinin06}.}             
\label{table:1}      
\centering                          
\begin{tabular}{c c c c c c c c}        
\hline\hline                 
$Log(M_{200})$ & $R_{200}$ & $n_0$ & $r_c/R_{200}$ & $r_s/R_{200}$ & $\alpha$ & $\beta$ & $\epsilon$ \\
$M_{\odot}$ & [Mpc] & $[10^{-4}cm^{-3]}$ & & & &  \\
\hline
12.43$\pm$0.43 & 250$\pm$35 & 1.38$\pm$0.03 & 0.10$\pm$0.01 & 2.15$\pm$0.03 & 0.89$\pm$0.1 & 0.39$\pm$0.06 & 2.86\\
12.70$\pm$0.45 & 390$\pm$46 & 1.11$\pm$0.03 & 0.12$\pm$0.01 & 2.05$\pm$0.04 & 0.72$\pm$0.2 & 0.30$\pm$0.07 & 2.86 \\
13.04$\pm$0.43 & 455$\pm$65 & 3.53$\pm$0.05 & 0.05$\pm$0.02 & 2.0$\pm$0.03 & 0.89$\pm$0.2 & 0.37$\pm$0.05 & 2.86\\
13.32$\pm$0.43 & 700$\pm$63 & 4.54$\pm$0.03 & 0.06$\pm$0.01 & 2.05$\pm$0.04 & 0.89$\pm$0.3 & 0.43$\pm$0.1 & 2.86 \\
13.55$\pm$0.46 & 780$\pm$71 & 9.25$\pm$0.04 & 0.04$\pm$0.02 & 1.95$\pm$0.03 & 0.8$\pm$0.2 & 0.43$\pm$0.07 & 2.86 \\
13.83$\pm$0.45 & 850$\pm$65 & 6.87$\pm$0.05 & 0.05$\pm$0.01 & 1.95$\pm$0.02 & 0.75$\pm$0.1 & 0.36$\pm$0.05 & 2.86 \\
14.15$\pm$0.41 & 1997$\pm$41 & 1.17$\pm$0.01 & 0.042$\pm$0.005 & 1.7$\pm$0.01 & 1.07$\pm$0.03 & 0.40$\pm$0.03 & 8.96$\pm$0.13 \\

\hline  \hline                                 
\end{tabular}
\end{table*}

\subsection{The average X-ray surface brightness profile of groups}\label{combine}
To derive gas density profiles for galaxy groups, it is crucial to obtain accurate X-ray surface brightness profiles. For this purpose, we use the average surface brightness profiles from \citet{Popesso2024c}, which have been rigorously validated using synthetic datasets that replicate the observed eROSITA X-ray and GAMA optical data, generated from the Magneticum simulations’ lightcones. These profiles are based on stacking GAMA groups limited to $z < 0.2$ within the eFEDS area.

The redshift limit ensures high completeness for the GAMA group sample down to $M_\mathrm{halo} \sim 10^{12.2} M_{\odot}$. The use of a halo mass proxy based on total group luminosity rather than velocity dispersion, makes the richness cut applied in \cite{pp24} unnecessary. As shown in \citet{Marini24b}, this luminosity-based proxy offers more reliable halo mass estimates and extends the completeness down to lower masses. According to \citet{Marini24b}, the mock GAMA group sample retains over 90\% completeness down to $\log(M_{200}/M_{\odot}) \sim 13.5$, decreasing to 80\% at $\log(M_{200}/M_{\odot}) \sim 13$, and to 65\% at $\log(M_{200}/M_{\odot}) \sim 12$. Contamination remains below 10\% for $\log(M_{200}/M_{\odot}) \gtrsim 13$, increasing to 20\% at lower masses.

In \citet{Popesso2024c} we apply the stacking procedure of \citetalias{pp24} by using the GAMA sample as prior catalog for the stacking in the eFEDS data. We consider six halo mass bins from $M_{200} > 3\times10^{12}$ $M_{\odot}$ to $10^{14}$ $M_{\odot}$, from Milky Way like groups to poor clusters. All priors in the halo mass bins are considered without distinction between detected and undetected groups, to measure the average X-ray surface brightness profile of the underlying group population. Only groups with a companion or a point source within $2\, R_{200}$ are discarded from the prior catalog because they would contaminate the background subtraction and the resulting average profile. Briefly, the stacking is done by averaging the background subtracted surface brightness profiles within the same annuli around the group center. The background is measured in a region between 2 to 3 Mpc from the group center. All events flagged as point sources in each annulus are excluded and the corresponding annulus area is corrected for the excluded point source area. All groups containing a point source or contaminated by close neighbors within $2\, r_{200}$ are excluded from the prior sample for the stacking. The X-ray luminosity from the stacked signal is derived in the 0.5-2 keV band by selecting only events with a rest frame energy in the selected band at the median redshift of the prior sample. The spectroscopic information is retained and corrected for the effective area to ensure an accurate estimate of the X-ray luminosity. We refer to \citetalias{pp24} and \citet{Popesso2024c} for a more detailed description of the procedure, including the AGN and XRB contamination based on the Magneticum model. The results of this analysis are shown in Fig. 2 of \citet{Popesso2024b}, which presents the average X-ray surface brightness profiles obtained in this way. 

Additionally, we include the average profile of detected clusters with $M_{200} > 10^{14} M_{\odot}$ at $z < 0.2$ from eFEDS, from \citet{LiuAng2022}. By cross-referencing the eROSITA eFEDS sample with optically selected groups, we identify an optical counterpart for each X-ray detection, except in two cases where associations were ambiguous. In one case, multiple lower-mass groups at the same redshift overlapped the region of the X-ray detection, and in the second case, the redshift provided by \citet{Klein2022} did not match that of the corresponding GAMA group. These two cases were excluded from our analysis. No optically selected groups remain undetected above $M_{200} > 10^{14} M_{\odot}$, confirming a consistent optical and X-ray selection of groups at high halo masses. As shown in \citetalias{pp24}, the stacked profiles of detected clusters are in agreement with the average profiles from \citet{LiuAng2022}, ensuring consistency throughout our study. The X-ray detected GAMA systems in eFEDS complement the stacked profiles in the $10^{14} < M_{200} < 10^{14.3} M_{\odot}$ range. For detected systems, we use the total optical luminosity as a mass proxy, maintaining the same selection function across the entire range analyzed. In Fig. \ref{a1} we show the surface brightness profiles for the groups in the seven halo mass bins, the first six obtained by stacking and the last one by averaging.

To complement our measure at higher halo masses, we use the average $f_{gas}$ profile of the CHEX-MATE cluster sub-sample used for the stack of eROSITA data by \citet{Lyskova23}. The CHEX-MATE cluster sample \citep{CHEX-MATE21} is an unbiased, signal-to-noise limited sample of 118 galaxy clusters detected by Planck via the Sunyaev-Zel'dovich effect. It is composed of clusters at
$ z < 0.2$ with masses $2\times 10^{14} < M_{500}/M_{\odot} < 9\times 10^{14}$ from the PSZ2 catalogue \citep{Planck2016}. The CHEX-MATE subsample used for the stacking comprises 38 systems in regions with minimal background variations due to proximity to the Galactic plane and the Cygnus-X star formation region, and relatively far from the North Polar Spur. We derive the mean $M_{200}$ and $R_{200}$ from the average $M_{500}$ and $R_{500}$ estimates listed in \citet{Lyskova23}, by assuming a NFW profile with concentration of 6 \citep[see also][]{DM14}.

\section{The gas mass estimate}\label{fgas}

We estimate the gas mass profile as described in \cite{LiuAng2022}. Specifically, we use a \citet{Vikhlinin06} electron number density model 

\begin{figure}
\includegraphics[width=0.5\textwidth]{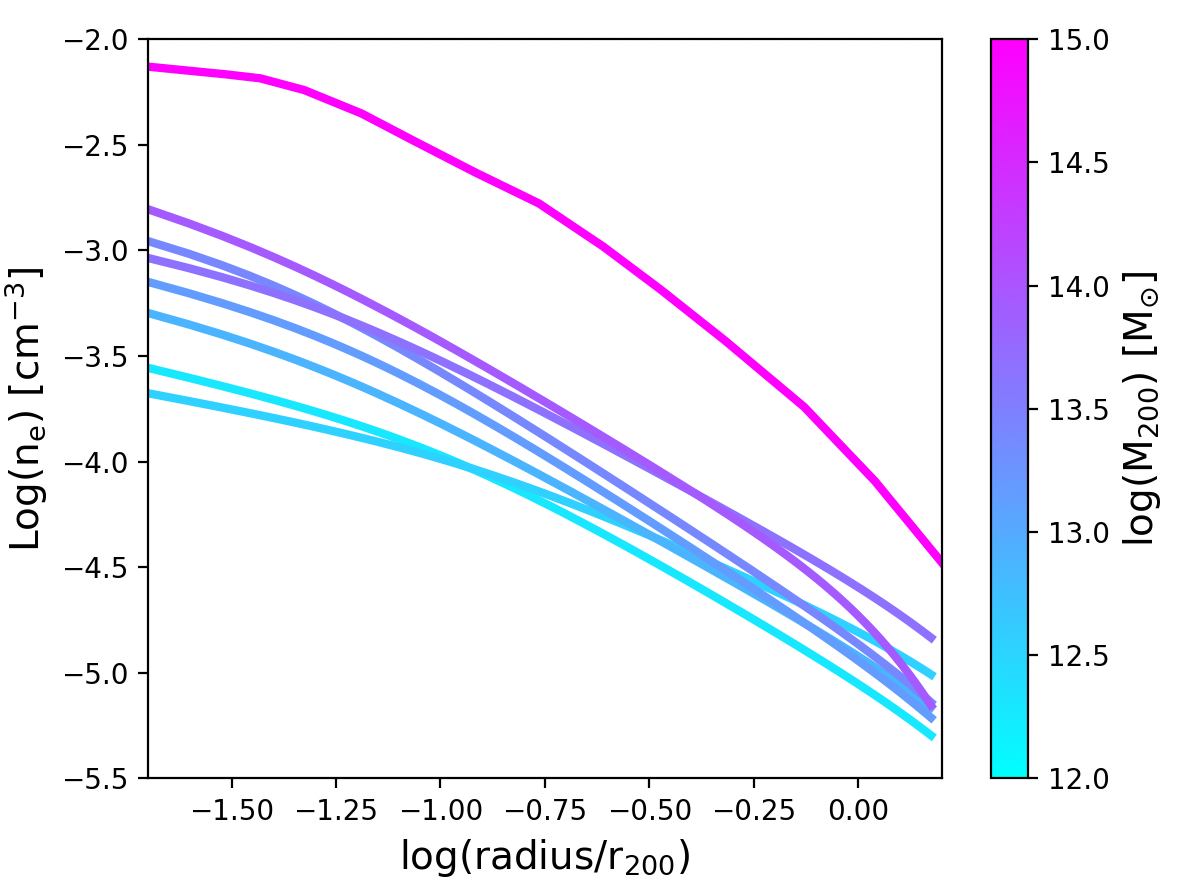}
\caption{Electron density profiles of groups and clusters in the halo mass bins studied here. The profiles are color-coded as a function of $M_{200}$ as shown in the color bar. The magenta profile is obtained from the stacking in eROSITA of the CHEXMATE clusters and it is taken from \citet{Lyskova23}.}
\label{fig1}
\end{figure}

\begin{equation}
n_{\mathrm e}^2(r) = n_0^2 \cdot \left( \frac{r}{r_{\mathrm c}} \right)^{-\alpha} \cdot \left( 1 + \left( \frac{r}{r_{\mathrm c}} \right)^2 \right)^{-3\beta+\alpha/2} \cdot \left(  1 + \left( \frac{r}{r_{\mathrm s}} \right)^3 \right)^{-\epsilon/3},
\end{equation}

\noindent where $n_0$ is the normalization factor, $r_{\mathrm c}$ and $r_{\mathrm s}$ are the core and scale radii, $\beta$ controls the overall slope of the density profile, $\alpha$ controls the slope in the core and at intermediate radii, and $\epsilon$ controls the change of slope at large radii. This is used to estimate and integrate along the line of sight the X-ray emissivity,  $n_e(r)^2\Lambda(kT, Z)$ profile, where $\Lambda(kT, Z)$ is the cooling function depending on the gas temperature and metallicity. The cooling function is derived by assuming a gas temperature from the $M-T_X$ relation of \cite{Lovisari2015} at the mean $M_{500}$ of the halo mass bin. As in \citet{LiuAng2022}, we assumed a metallicity of the ICM of $0.3\, Z_{\odot}$, adopting the solar abundance table of \citet{asplund09}, that includes the He abundance. This assumption is consistent with the average metallicity estimated in poor cluster cores by \citet{lovisari19} and with Magneticum predictions, although the observed scatter is large. The error due to these assumptions is estimated in \citep[][see also next session]{Popesso2024b} based on the Magneticum mock observations. The effect on X-ray emissivity depends mainly on the systematics of the halo mass proxy. For the GAMA group sample, the use of the total luminosity proxy results in a scatter of 40\% in the emissivity estimate at fixed metallicity. The assumption of a fixed metallicity of $0.3 \, Z_{\odot}$ leads to a variation in emissivity of 37\%. Since these effects cannot be corrected, they are included in the error budget and summed in quadrature with the statistical errors.

The projected number density model is convolved with the eROSITA PSF and fitted to the data. Differently from \cite{LiuAng2022} we leave all parameters free in the fit, but $\epsilon$, which rules the change of slope at large radii. This is poorly constrained due to the relatively low SNR in the outskirts region. Thus, we fixed the value to the cluster value provided by \citet{Vikhlinin06}. Only for the average profile of the detected poor clusters at $14 < \log(M_{200}/M_{\odot}) < 14.3$, the SNR is high enough to constrain the change of slope at large radii. The best-fit line to the X-ray surface brightness profiles and the best-fit parameters are reported in Fig. \ref{a1}. The electron density profiles obtained in this way are shown in Fig. \ref{fig1}. We include also the profiles of \citet{Lyskova23} for the CHEX-MATE clusters.

We compute the ICM mass of a system within a given aperture, by using the best-fit model for the electron density  profile,
\begin{equation}
M_{\mathrm{ICM}} = 4\pi \mu_\mathrm{e} m_\mathrm{p}\int_{0}^{R} n_\mathrm{e}(r)\ r^{2}\ {\rm d}r,
\end{equation}
\label{eq_gas}

where the average nuclear charge and mass are $A\sim1.4$ and $Z\sim1.2$, and $\mu_{\rm e}=A/Z\sim1.17$. 

\subsection{Validation of the stacking procedure and gas mass estimate}

The stacking procedure applied to eROSITA data in different mass bins, as validated by \citet{Popesso2024b}, employs a mock dataset generated from the L30 lightcone of the Magneticum simulation. This dataset incorporates mock eROSITA observations down to the eRASS:4 depth (corresponding to two years of observations) and a GAMA-like spectroscopic galaxy survey. A Friends-of-Friends (FoF) algorithm, as described by \citet{Robotham2011}, is applied to the mock galaxy data to generate an optical group catalog resembling the GAMA-selected sample. The performance of this FoF algorithm, including completeness and contamination, is extensively analyzed in \citet{Marini24b}.

\citet{Popesso2024b} replicate the analysis framework of \citet{Popesso2024c} by using this mock dataset to test the reliability of the stacking method and gas mass estimations. Key elements of the mock analysis—halo mass limits, mass proxies, and bin sizes—are matched to those in the real dataset, ensuring a robust validation of the procedure.

The study addresses key systematic uncertainties, including selection biases in optical catalogs, mis-centering between optical and X-ray datasets, AGN and XRB contamination, and scatter in the halo mass proxy. Notably, completeness and contamination in optical catalogs were shown to be sufficient to avoid significant biases, while mis-centering is mitigated due to small positional offsets compared to the eROSITA PSF. AGN and XRB contamination, prominent in low-mass halos ($M_{200} < 10^{13} M_{\odot}$), are effectively modeled and subtracted, ensuring minimal impact on X-ray surface brightness estimates.

The stacking analysis demonstrates excellent agreement between the input and retrieved X-ray surface brightness profiles. Low-mass halos exhibit a maximum residual overestimation of 30\%, attributable to contamination in the halo mass proxy bins. In higher mass halos, profiles are accurately reproduced. Across all mass bins, the stacking procedure proves robust, confirming no significant biases in the analyzed halo mass range, even when only 60\% of the population is captured by the detection algorithm.

Regarding gas mass estimations, the study assesses the systematic impact of inferring gas temperatures from the $T_X$-mass scaling relation and assuming a fixed metallicity of $0.3Z_{\odot}$. For the R11 algorithm, the analysis shows a 40\% scatter in the retrieved emissivity values. By incorporating uncertainties in gas temperature and metallicity into the error budget, the stacking method reliably reconstructs input electron density profiles, further validating its robustness \citep[see Fig. 13 of][]{Popesso2024b}. This approach ensures that systematic uncertainties are accounted for when interpreting the results from real observations.

\begin{figure}
\includegraphics[width=0.5\textwidth]{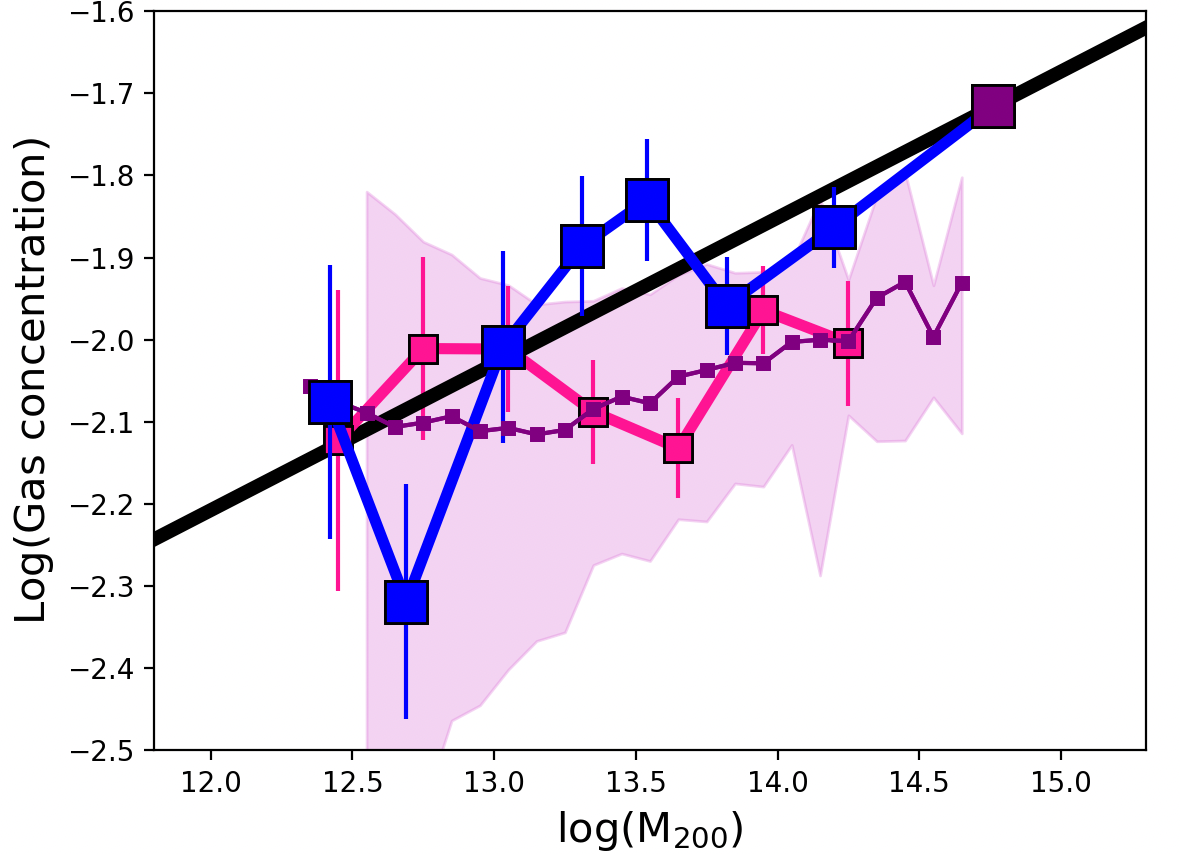}
\caption{Relation between gas concentration, defined as the ratio of hot gas mass within $0.1 \, R_{200}$ to that within $R_{200}$, versus halo mass. The blue points represent results derived from the stacked data. The big purple point shows the result based on the electron density profile of the CHEX-MATE clusters stacked in eROSITA \citep{Lyskova23}. The solid black line represents the best-fit relation based on the combined stacks of eFEDS and CHEX-MATE data. The small purple points indicate the mean relation predicted by the Magneticum simulation, with the shaded region illustrating the dispersion around this relation. The pink squares display the results obtained using the same stacking technique applied in this study to the X-ray surface brightness profiles derived from mock galaxy groups selected using the \citet{Robotham2011} algorithm on the mock eROSITA data from the L30 lightcone of Magneticum.}
\label{cgas}
\end{figure}

\section{Results}\label{res}

\subsection{The gas concentration}

To quantify how the shape of the gas profiles varies with system mass, we estimate the gas concentration ($c_{gas}$) as the ratio of the gas mass within $0.1 \, R_{200}$ to that within $R_{200}$. Figure \ref{cgas} shows the variation of gas concentration with halo mass. A clear trend is observed, with gas concentration increasing with mass; specifically, the CHEX-MATE clusters are approximately 2.5 times as concentrated as Milky Way-sized groups, following a dependence of $c_{gas} \propto M_{200}^{0.178\pm0.09}$. This factor is 3 times larger than errors deriving from observational error of the stacking, including the uncertainty due to the estimate of the gas emissivity by assuming the gas temperature and metallicity, as described in the previous section. Additionally, according to the standard hierarchical formation paradigm, dark matter concentration ($c_{200}$) decreases monotonically with mass \citep[see also][]{DM14}. The opposite trend observed for gas concentration suggests the influence of non-gravitational processes, most likely AGN feedback.

We propose that the observed increase in gas concentration from low-mass groups to massive clusters results from AGN feedback evacuating gas more effectively in low-mass halos due to their shallower potential wells compared to clusters. To explore this trend further, we examine the L30 lightcone of the Magneticum simulation, finding a positive correlation with a Spearman coefficient of 0.64 and a probability of no correlation of $\sim10^{-5}$. However, Magneticum predicts a flatter dependence of $c_{gas} \propto M_{200}^{0.090\pm 0.01}$.

Figure \ref{cgas} also presents $c_{gas}$ values obtained by applying the same analysis as in this study to the stacked surface brightness profiles of the mock dataset analogous to the one used here \citep{Marinia,Popesso2024b}.  Figure \ref{cgas} shows that our approach, including uncertainties and systematics of the group selection, halo mass proxy choice, and stacking technique, accurately reproduces the $c_{gas}-M_{200}$ relation predicted by Magneticum within $1\sigma$. This consistency suggests that the results obtained from the observed stacked profiles are robust.

\begin{figure}
\includegraphics[width=0.5\textwidth]{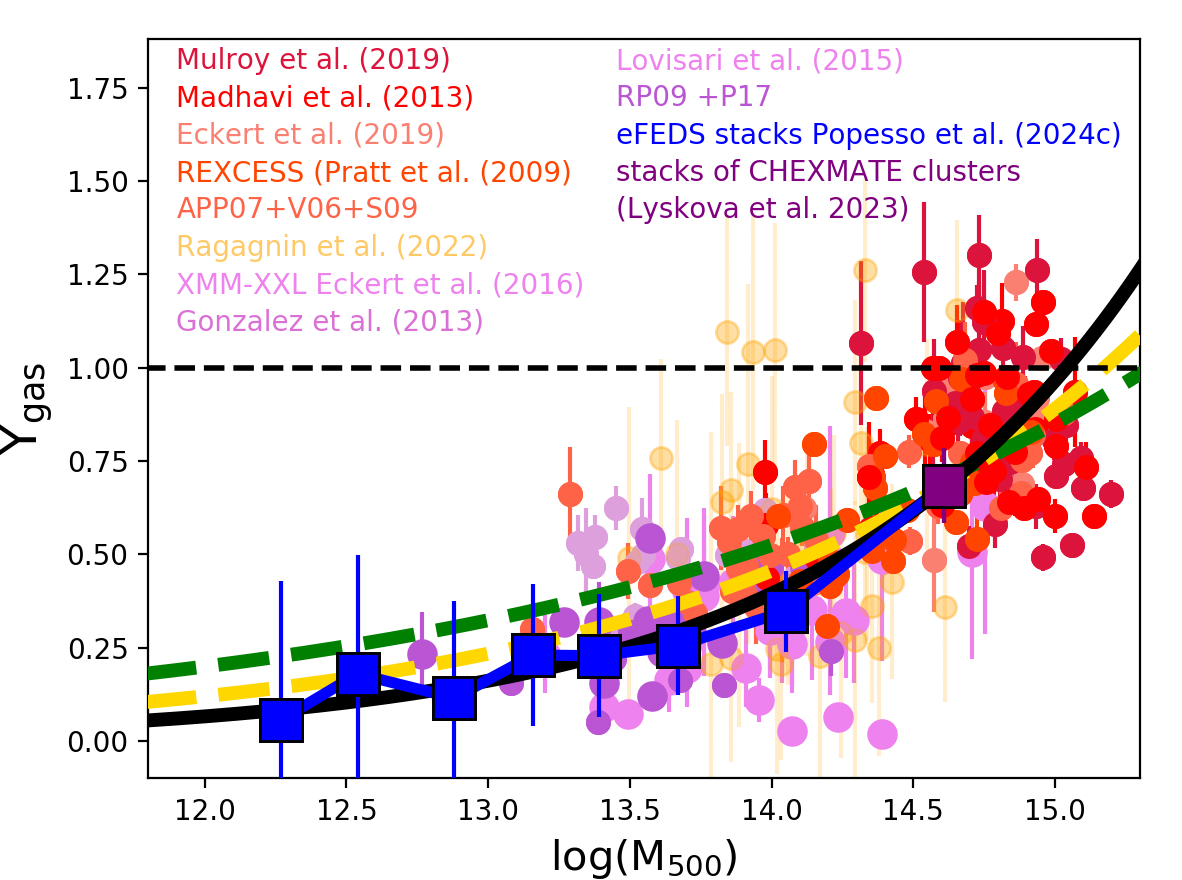}
\caption{$Y_\mathrm{gas}$-$M_{500}$ relation. The filled circles represent a compilation of literature data, color-coded as a function of the sample reference as shown in the picture. The blue squares indicate the $Y_\mathrm{gas}$ derived from the stacked groups of Popesso et al. (2024c). The purple square indicates the value derived from the CHEXMATE clusters stacked in eROSITA data by \citet{Lyskova23}. The solid black curve represents our best fit to stacked points including the CHEXMATE clusters. The dashed yellow and green curves represent the best fit of \citet{pratt09} and \citet{Eckert16}, respectively. The $Y_\mathrm{gas}=1$ corresponding to the \citet{planck18} value is indicated by the black dashed line.}
\label{fgas_m500}
\end{figure}

\begin{figure}
\includegraphics[width=0.5\textwidth]{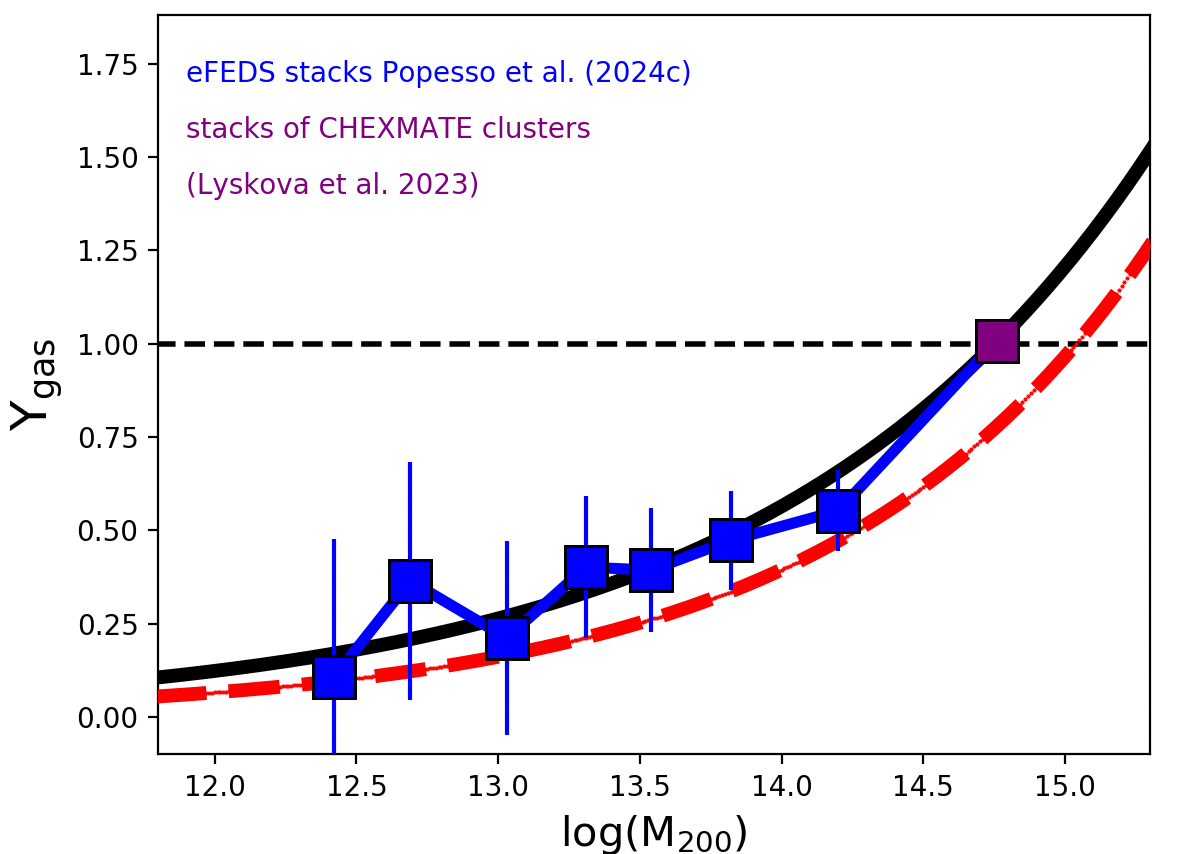}
\caption{$Y_\mathrm{gas}$-$M_{200}$ relation. The blue squares indicate the $Y_\mathrm{gas}$ derived from the stacked groups of \citet{Popesso2024c} The purple square indicates the value derived from the CHEXMATE clusters stacked in eROSITA data by \citet{Lyskova23}. The solid black curve represents our best fit to stacked points including the CHEXMATE clusters. The red dashed line shows the best fit of the $Y_\mathrm{gas}-M_{500}$ relation of Fig. \ref{fgas_m500}.}
\label{fgas_m200}
\end{figure}

\subsection{The observed $Y_\mathrm{gas}-M_\mathrm{halo}$ relation}\label{ygas}
We define the fraction of gas mass within a radius $r$ as:
\begin{equation}
f_{gas}(<r)=M_\mathrm{gas}(<r)/M_\mathrm{tot}(<r)
\end{equation}
where $M_\mathrm{gas}$ is estimated from eq.~(\ref{eq_gas}) and $M_\mathrm{tot}(<r)$ is the total mass contained within the radius $r$. In this analysis, we study the gas fraction contained within $R_{500}$ and $R_{200}$. We normalize $f_{gas}$ to the universal baryon fraction \citep[$\Omega_{b}/\Omega_{m}=0.154$,][]{planck18} to obtain $Y_\mathrm{gas}(<r)$. 

Fig. \ref{fgas_m500} shows $Y_\mathrm{gas}$ estimated within $R_{500}$ versus $M_{500}$. 
$Y_\mathrm{gas}$ varies by a factor of $\sim 10$ from $M_{500} \sim 5\times 10^{12}$ $M_{\odot}$, Milky Way-sized groups, to $M_{500} \sim 5\times 10^{14}$ $M_{\odot}$, the CHEXMATE clusters. The same is observed for the $Y_\mathrm{gas}$ estimated within $R_{200}$ (see Fif. \ref{fgas_m200}) The best-fit power-law relation to the data including the CHEXMATE stacked point is:
\begin{equation}
    f_\mathrm{gas,500}=2.23\pm{0.18}\times10^{-7} (M_{500}/M_{\odot})^{0.39\pm0.02}
    \label{eq_fgas500}
\end{equation}

\begin{figure*}
\includegraphics[width=\textwidth]{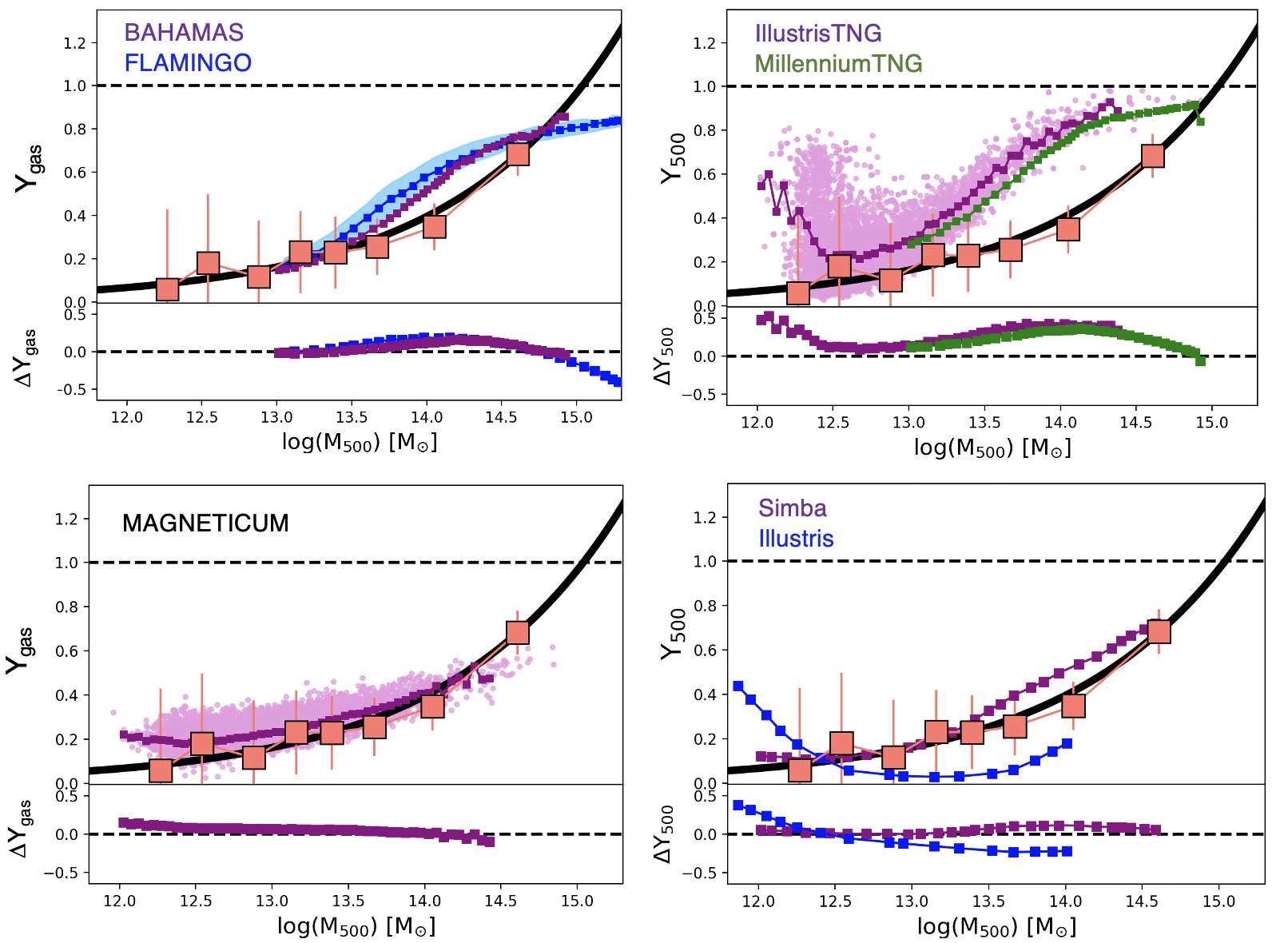}
\caption{Comparison of the observed $Y_\mathrm{gas}$ versus mass relations within $R_{500}$ with different hydrodynamical simulations. The color code for each dataset is indicated in each panel. The light orange squares show the stacked points of Fig. \ref{fgas_m500}, while the black line shows our best-fit relation. The bottom subpanels show the residuals of each dataset to our best fit. These are expressed in dex ($log(f_{gas})-log(fit)$). {\it{Top-left panel:}} Comparison of $Y_{gas}$-$M_{500}$ relation with the predictions of the BAHAMAS \citep{Salcido23}, and FLAMINGO \citep{Schaye23} simulations. The blue shaded region indicates the $1\sigma$ uncertainty of the relation in FLAMINGO as reported in the corresponding paper. These two simulations use similar galaxy evolution models and feedback implementations.  {\it{Top-right panel:}} Comparison of $Y_{gas}$-$M_{500}$ relation with the predictions of the IllustrisTNG \citep{pillepich19} and of MillenniumTNG \citep{Pakmor22} simulations. For IllustrisTNG we show the individual estimates (pink points) and the mean relation (purple squares). For MillenniumTNG we report the mean relation (green squares) retrieved in the corresponding paper. These two simulations use similar galaxy evolution models and feedback implementations. {\it{Bottom-left panel:}} Comparison of $Y_{gas}$-$M_{500}$ relation with the predictions of the {\it Magneticum} \citep{Dolag16}. The pink points indicate the individual estimates, while the purple squares indicate the mean relation. {\it{Bottom-right panel:}} Comparison of $Y_{gas}$-$M_{500}$ relation with the predictions of the {\it Simba} \citep{Dave19} and Illusris \citep{Genel2014}.}
\label{fig2}
\label{fsims}
\end{figure*}

Our best fit of eq.~(\ref{eq_fgas500}) is consistent within $1\sigma$ with the estimates of \citet{pratt09} and \citet{Eckert16}. Nevertheless, it has a similar slope but a lower normalization in comparison to the estimates of \citet{Sun2009}, \citet{Lovisari2015}, and \citet{Eckert21}. These estimates are based on ROSAT galaxy group samples, that include only the brightest groups of the local Universe due to the high RASS flux limit. The stacked data, instead, are representative of the bulk of the underlying halo mass population as tested in \citet{Popesso2024b} through the analysis of the analog mock dataset generated from Magneticum. Thus, they include also groups and poor clusters whose gas distribution might have been heavily affected by non-gravitational processes such as AGN feedback. 

In addition to the combination of the eFEDS stacks and the CHEX-MATE stacked point, we also show in Fig. \ref{fgas_m500} a comprehensive compilation of literature data providing the same estimate for $z < 0.2$ systems. These are the cluster sample of \citet{Mulroy19}, \citet{Madhavi13}, \citet{eckert19}, the REXCESS sample of \citet{pratt09}, the poor cluster and group sample of \citet{Arnaud07}, \citet{Vikhlinin06}, and \citet{Sun2009} used in \citet{pratt09} (APP05$+$V09$+$S09 in Fig.\ref{fgas_m500}), the XMM-XXL sample of \citet{Eckert16}, the group samples of \citet{Gonzalez13} and \citet{Lovisari2015}, the poor cluster sample of \citet{Ragagnin22}, and the group sample of \citet{Rasmussen09} and \citet{Pearson17} (RP09$+$P17 in Fig.\ref{fgas_m500}). We transformed all the values to our adopted cosmology. We compare the distribution of the literature compilation to our best fit. Given the very different selection functions of the collected data, it is not surprising that they scatter largely around our best fit. Nevertheless, we find consistency in the distribution of the observed systems and the best fit provided here and based on the stacked profiles. In particular, we point out that the XMM-XXL groups of \citet{eckert19}, which are selected at a flux limit similar to eFEDS, exhibit $f_{gas}$ values consistent with the mean values of the eFEDS detections and stacks at $M_{500}\sim 10^{13.5}- 10^{14}$ $M_{\odot}$. 

We also present here in Fig. \ref{fgas_m200}, for the very first time, the $f_{gas}$ versus halo mass relation, estimated within $R_{200}$. Until recently, measuring the $f_{gas}$ beyond $R_{500}$ was hampered by the lack of sensitivity of previous instruments. The stacking analysis can overcome this problem, as shown by \citetalias{pp24} and \citet{Lyskova23}. The combination of eFEDS stacks and the CHEX-MATE stack allows us to determine $f_{gas}$ over the largest halo mass range, and for the largest radius, ever probed so far in groups and clusters. The best-fit relation is:
\begin{equation}
    f_\mathrm{gas,200}=2.09\pm{0.14}\times10^{-6} (M_{200}/M_{\odot})^{0.33\pm0.02}
    \label{eq_fgas200}
\end{equation}
At the Milky Way-sized group scale, $f_{gas}$ is $\sim$20-40\% of the cosmic value. It goes above 50\% for massive groups and poor clusters and it reaches the cosmic value at the CHEX-MATE cluster mass scale. The slope of best fit of eqs.~(\ref{eq_fgas500}) and (\ref{eq_fgas200}) are consistent within $2\sigma$, but the fraction of gas nearly doubles from $R_{500}$ to $R_{200}$. 

\subsection{Comparison with simulations}\label{sims}
We compare the results of our analysis with the predictions of several hydrodynamical simulations, each implementing a different BH feedback model. We consider, in particular, the predictions of MillenniumTNG \citep{Pakmor22}, FLAMINGO \citep{Schaye23}, BAHAMAS \citep{McCarthy17}, SIMBA \citep{Dave19}, Illustris \citep{Genel2014}, IllustrisTNG \citep{pillepich19} and {\it Magneticum} \citep{Dolag16}. 

For MillenniumTNG, FLAMINGO, BAHAMAS, Illustris, and SIMBA, the $f_{gas}$-$M_{500}$ relation is provided in the reference paper. We normalize such relation to the value of $\Omega_b/\Omega_m$ according to the cosmology implemented in each simulation.
Only for the IllustrisTNG and Magneticum, we derive the $f_{gas}-M_{halo}$ relation ourselves. For IllustrisTNG, we select halos in TNG-300, corresponding to the largest cosmological box available, including 75000$^3$ particles in a 300 h$^{-1}$ cMpc$^{3}$ volume. The dark matter (initial gas) particles have mass $5.9\times 10^{7}$ h$^{-1} M_{\odot}$ ($1.1\times 10^{7}$ h$^{-1} M_{\odot}$) and Plummer equivalent softening $0.74$ kpc ($0.19$ ckpc). Details on the simulation and the astrophysical subgrid models implemented can be found in \citet{pillepich19}. 
We select halos at redshift $z=0$ since we find no evolution of the $Y_\mathrm{gas}-M_{halo}$ relation in the redshift range $z=0$-0.2 (i.e., the redshift range of the surveys considered here). As for {\it{Magneticum}}, halos are drawn from {\it{$Box2b/hr$}} which includes a total of 2$\times$ 2880$^{3}$ particles in a volume of (640 h$^{-1}$ cMpc)$^{3}$. The dark matter (initial gas) particles have mass $6.9\times 10^{8}$ h$^{-1} M_{\odot}$ ($1.4\times 10^{8}$ h$^{-1} M_{\odot}$) and Plummer equivalent softening $3.75$ h$^{-1}$ kpc ($3.75$ h$^{-1}$ kpc). We select halos in the last snapshot available at redshift $z=0.25$. To further prove the consistency of our investigation, we repeat the analysis with $Box2/hr$ -- $2\times1584^3$ particles in a (352 h$^{-1}$ cMpc)$^{3}$ volume -- both at $z=0$ and $z=0.3$ to rule out possible intrinsic redshift evolution of such relation. More details on the simulations and the astrophysical subgrid models implemented can be found in \cite{Dolag16}. 
In both simulations, we gather a representative sample of clusters and group-size halos by randomly selecting $10^4$ halos with log~$M_{500}/M_{\odot} \geq$ 12.2 in each simulation. For each of them, we compute the baryon fraction $Y_\mathrm{gas}$ within $R_{500}$ and $R_{200}$ considering all the hot gas particles (i.e., $T > 10^6$ K).

The result of the comparison is shown in Fig. \ref{fsims}. Compared to our observational results, $Y_\mathrm{gas}$ is over-predicted at given $M_{halo}$ by MillenniumTNG and IllustrisTNG, which apply the same galaxy physical model \citep[see also][for an alternative estimate in IllustrisTNG300]{pop22}. FLAMINGO, and BAHAMAS, which have similar approaches for the BH feedback implementation, are closer to our observed relation, although they do not follow the observed power law, in particular in the $M_{halo}$ range $10^{13.5-14.5}$ $M_{\odot}$ \citep[see also][]{Salcido23}. The {\it Magneticum} simulation reproduces the observed relation reasonably well, with a slightly higher normalization that remains consistent within $1\sigma$ of the observations \citep[see also][]{Angelinelli22}. SIMBA also shows agreement within $1\sigma$, albeit with slightly lower concordance. In contrast, Illustris exhibits an overly efficient evacuation of gas from halos across all masses. It is important to note that \citet{Genel2014} provides the gas fraction relation for all phases of gas rather than exclusively the hot phase. Consequently, the relation depicted in the bottom-right panel of Fig. \ref{sims} represents an upper limit to the hot gas-only relation.

\begin{figure}
\includegraphics[width=0.5\textwidth]{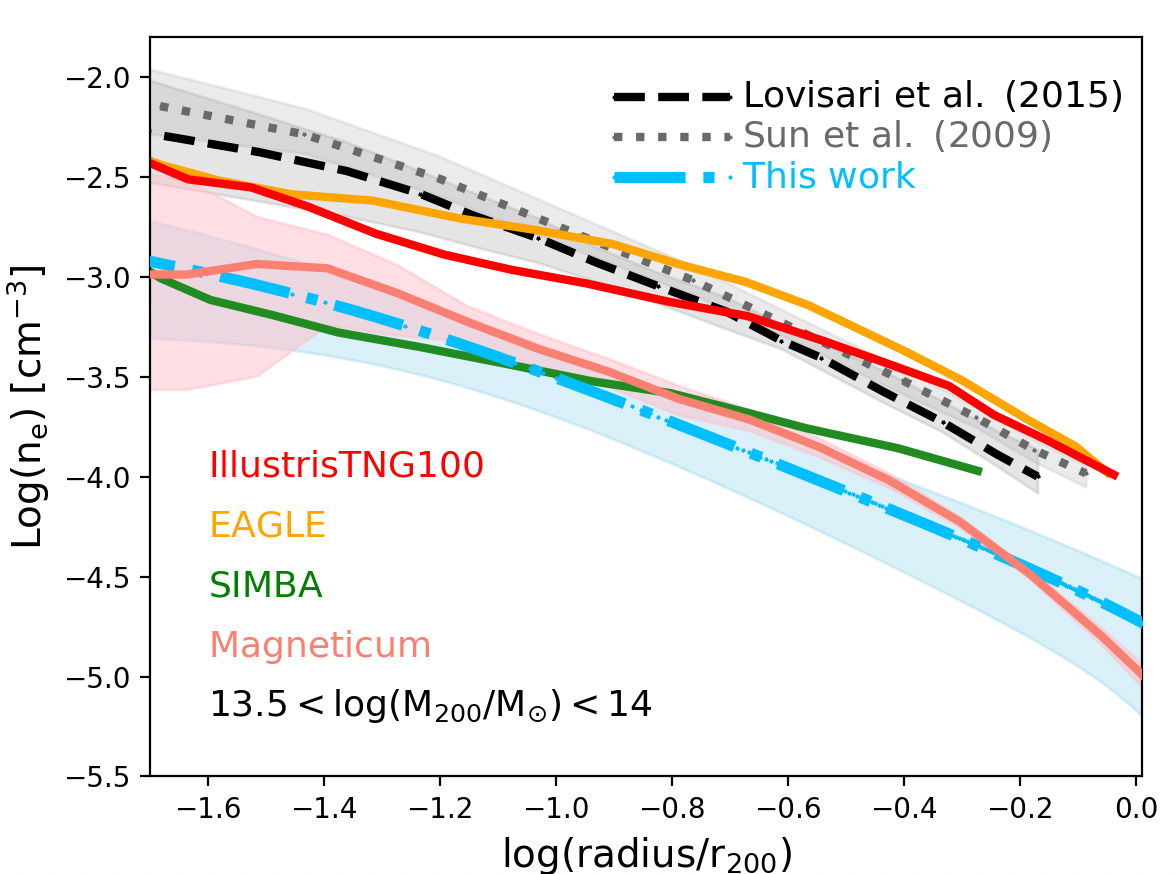}
\caption{Electron density profile for groups at $10^{13} M_{\odot} < M_{200} < 10^{14} M_{\odot}$ from this work (blue dashed-dotted curve), \citet{Lovisari2015} (black dashed curve) and \citet{Sun2009} (grey dotted curve), compared with the predictions of the hydrodynamical simulations in the same $M_{200}$ mass bin: {\it{Magneticum}} (pink curve), IllustrisTNG100 (red curve), EAGLE (orange curve) and Simba (green curve). With the exclusion of the Magneticum data provided by \citet{Popesso2024b}, the profiles of the simulations are taken from the review of \citet {Oppenheimer21}. }
\label{ne_sim_profile}
\end{figure}

The mass range of $10^{13.5}-10^{14} M_\odot$, where most predictions deviate most significantly from the observed relation, warrants further investigation. To this end, we compare the electron density profiles derived from our stacking analysis to predictions from simulations. For {\it Magneticum}, the electron density profile is obtained directly from simulation data, while other profiles are taken from the review by \citet{Oppenheimer21}. As shown in Fig. \ref{ne_sim_profile}, the profiles from {\it Magneticum} and SIMBA align most closely with the observations. In contrast, the normalizations of the EAGLE \citep{Schaye15} and IllustrisTNG simulations exceed observational results by a factor of approximately three. These results are more consistent with those from studies of bright X-ray groups, which naturally contain a higher gas fraction \citep[e.g.,][]{Lovisari2015, Sun2009}.

The discrepancies among simulations are largely attributable to differences in the treatment of galaxy formation physics and the associated timing and effects of black hole feedback on the intracluster medium (ICM). These variations make it challenging to isolate specific drivers of the observed differences, as they likely arise from a combination of factors. For example, \citet{Kauffmann2019} provide a compelling illustration of this complexity through a trace particle analysis comparing Illustris \citep{Nelson2015} and IllustrisTNG. Their study demonstrates that Illustris evacuates hot gas from galaxy groups more effectively than IllustrisTNG \citep[see also][]{Hadzhiyska2024}. The findings by \citet{Kauffmann2019} attribute these differences to the subgrid models of feedback, which affect the timing of gas displacement, cooling, and collapse during galaxy evolution. However, the observed differences in the dependence on group or cluster mass suggest that an important factor may be the inclusion (or lack thereof) of halo mass-dependent feedback parameters. For instance, kinetic feedback models vary across simulations and are absent in {\it Magneticum}, which could play a critical role in shaping the observed trends.



\begin{figure}
\includegraphics[width=0.5\textwidth]{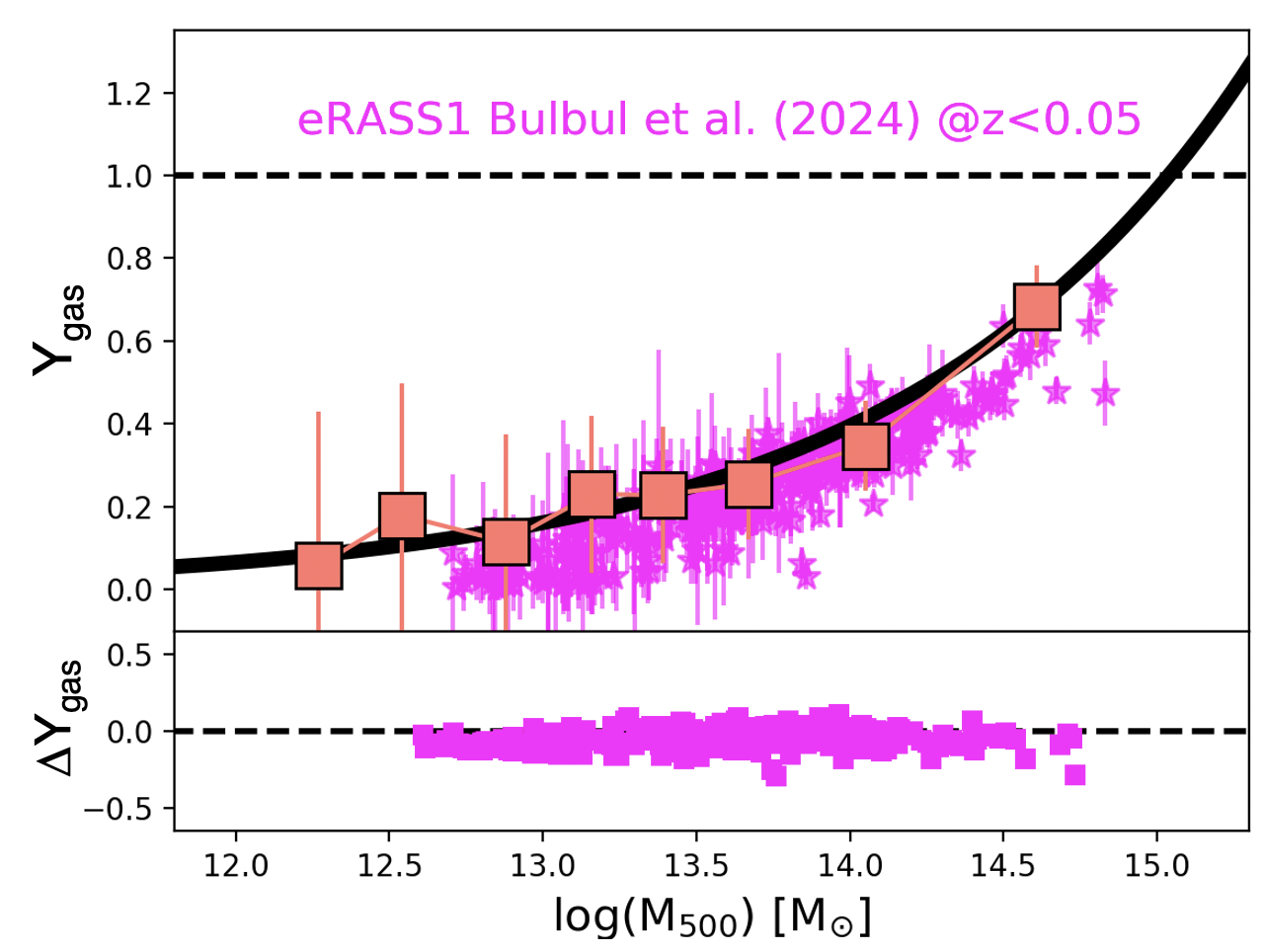}
\caption{{\it{Upper panel:}} Comparison of the observed $Y_\mathrm{gas}$ versus mass relations within $R_{500}$ with the $f_{gas}$ values within the same radius as given by \citet{Bulbul24} for the eRASS:1 extended objects at $z<0.05$ and with an optical counterpart. {\it{Lower panel:}} residuals of the eRASS:1 data from the best fit of Fig. \ref{fgas_m500} expressed in dex ($log(f_{gas})-log(fit)$.}
\label{bulbul}
\end{figure}

\subsection{Comparison with eRASS:1 data}
To further validate our findings, we compare them with a subsample of groups and clusters selected from eRASS:1 observations as presented by \citet{Bulbul24}. This subsample, illustrated in Fig. \ref{bulbul}, comprises all clusters and groups at $z < 0.05$ with a well-defined optical counterpart in the matched catalog provided in the eRASS:1 data release, as described by \citet{Bulbul24}. These systems include only those with more than five spectroscopic members. A flux cut of $10^{-13.5}$ erg s$^{-1}$ cm$^{-2}$ was applied to ensure a 50\% complete sample for systems with $M_{500} > 6 \times 10^{13}$ $M_{\odot}$, following the completeness criteria outlined by \citet{Bulbul24}. Additionally, Fig. \ref{bulbul} includes systems with masses below this completeness limit.

The gas masses reported by \citet{Bulbul24} were estimated using the methodology described in \citet{LiuAng2022}, which was previously applied to eFEDS-detected systems. Total masses were derived using scaling relations calibrated with weak lensing mass estimates \citep[see][for further details]{Bulbul24}. In these analyses, eROSITA data were simultaneously fitted to extract $M_{500}$ and temperature, iteratively refining the aperture size to define $R_{500}$ such that the data aligned with the observed $L_X-M_{500}$ scaling relation. As a result, all derived measurements are inherently correlated.

Despite these correlations, we find excellent agreement between our best-fit $f_{gas}-M_\mathrm{halo}$ relation and the independent estimates from \citet{Bulbul24}. This consistency further underscores the robustness of our results and strengthens confidence in the methodologies employed.

\section{Discussion \& Conclusions}\label{conc}
Using the stacking results from optically selected groups in eFEDS and the eROSITA stacked surface brightness profile of the CHEX-MATE clusters, we provide a comprehensive measurement of the $f_{gas}-M_\mathrm{halo}$ relation over the largest halo mass range and to the largest radii ever probed for galaxy systems at $z < 0.2$. Extensive tests conducted on an analogous mock dataset ensure that our results are robust against selection biases. We find that the $f_{gas}-M_\mathrm{halo}$ relation is well-described by a power law with a consistent shape (within $1-2\sigma$) when estimated within both $R_{500}$ and $R_{200}$ radii. This relation is in agreement with other studies that utilize X-ray survey data of similar depth, such as eFEDS and XXM-XXL \citep[see e.g.,][]{Eckert16}. However, our results indicate a lower $f_{gas}$ in the group regime compared to studies based on X-ray bright groups \citep[e.g.,][]{Sun2009, Lovisari2015, Eckert21}. We interpret this discrepancy as a selection effect, with X-ray brightest nearby groups occupying the upper envelope of the $f_{gas}-M_\mathrm{halo}$ relation at fixed halo mass.

Comparing our observed $f_{\mathrm{gas}}-M_{\mathrm{halo}}$ relation with predictions from various hydrodynamical simulations reveals that most current state-of-the-art models tend to over-predict the gas fraction across different halo mass ranges. Only {\it Magneticum} and, to a lesser extent, SIMBA align with the observed relation. Interestingly, the most significant discrepancies are not evident at the Milky Way group mass scale, where simulations generally perform well, but rather at the scale of massive groups and poor clusters ($13.5 < \log(M_{500}/M_{\odot}) < 14.3$). These discrepancies are unlikely to arise from systematic biases in our $f_{\mathrm{gas}}$ measurements, as our best-fit relation aligns well with literature data compilations (Fig. \ref{fgas_m500}). Instead, while {\it Magneticum} and SIMBA accurately reproduce the electron density profiles derived from the data, EAGLE and IllustrisTNG overestimate the normalization of these profiles, leading to an overprediction of gas mass fractions.

The low $f_{\mathrm{gas}}$ values observed at the group mass scale, approximately 20–40\% of the cosmic baryon fraction, suggest that a significant portion of the hot gas may reside beyond the virial region or exist in non-X-ray-emitting phases. Simulations predominantly support the first scenario, where feedback mechanisms expel gas to regions far beyond the virial radius, particularly in lower-mass halos. Studies such as \citet{Angelinelli22} and \citet{Reza2023} demonstrate that in both {\it Magneticum} and IllustrisTNG, most baryons in the form of hot gas are displaced well beyond the virial radius in Milky Way-sized halos and massive groups. This displacement implies that the halo closure radius—where baryon content aligns with the cosmic value—extends significantly beyond $R_{200}$ for most group-sized systems. However, in IllustrisTNG, this gas expulsion is insufficient at the massive group and poor cluster scales, where our observations reveal discrepancies in $f_{\mathrm{gas}}$ of up to a factor of three. This may indicate that the moderated or halo-mass-dependent feedback in IllustrisTNG, introduced to better match galaxy properties \citep{pillepich19}, does not expel gas as efficiently as the original Illustris model \citep{Nelson2015}.

This discrepancy is further highlighted in \citet{Hadzhiyska2024}, which reports kinematic Sunyaev-Zeldovich (kSZ) effect measurements from the Atacama Cosmology Telescope (ACT), stacked on the luminous red galaxy (LRG) sample of the Dark Energy Spectroscopic Instrument (DESI). These observations detect gas extending well beyond the virial radius at high significance ($> 40\sigma$), showing better agreement with Illustris predictions than with IllustrisTNG \citep[see also][]{amodeo21}. It is worth noting that the feedback adjustments in IllustrisTNG improve predicted galaxy properties, such as color and star formation rates, compared to Illustris \citep{Sparre2015, Donnari2021}. However, these adjustments appear to compromise the accuracy of gas distribution predictions within halos. Similarly, while {\it Magneticum} performs well in reproducing halo gas properties \citep[see also][]{Popesso2024b,Popesso2024c}, it overquenches local galaxy populations (Mazengo et al., in prep.), as does SIMBA \citep{Dave19}.

Other simulations, such as BAHAMAS and FLAMINGO, exhibit less extreme behavior compared to IllustrisTNG. While they do not perfectly reproduce the $f_{\mathrm{gas}}-M_{\mathrm{halo}}$ relation, the maximum disagreement is limited to a factor of two at $M_{500} \sim 10^{14} , M_{\odot}$. Additionally, their galaxy populations do not suffer from the strong overquenching of star formation observed in Magneticum, or SIMBA, although their lower resolution poses limitations compared to these simulations \citep{Schaye23,McCarthy17}.

We conclude that current hydrodynamical simulations face a critical challenge in balancing subgrid physics. While strong feedback mechanisms effectively expel gas from halos, they often suppress star formation too aggressively, leading to inconsistencies in modeling galaxy populations. The critical challenge lies not only in determining the magnitude of energy feedback required to deplete or prevent gas cooling and inflow but also in understanding when and where these processes occur. Addressing this requires robust observational constraints on halo gas and galaxy properties across cosmic time and varying environments, which remain largely lacking.

In conclusion, our estimates of $f_{gas}$ over a wide range of halo masses, from Milky Way-sized groups to clusters, and out to $R_{200}$ distances, set severe constraints to theoretical models that investigate the co-evolution of the hot intra-group medium in the presence of the central galaxy BH feedback, and highlights the importance of avoiding selection biases when deriving halo properties that can depend on the halo selection process itself.

\section*{Aknowledegements}
PP acknowledges financial support from the European Research Council (ERC) under the European Union’s Horizon Europe research and innovation programme ERC CoG CLEVeR (Grant agreement No. 101045437). AB acknowledges the financial contribution from the INAF mini-grant 1.05.12.04.01 {\it "The dynamics of clusters of galaxies from the projected phase-space distribution of cluster galaxies"}. KD acknowledges support by the COMPLEX project from the European Research Council (ERC) under the European Union’s Horizon 2020 research and innovation program grant agreement ERC-2019-AdG 882679. The calculations for the {\it Magneticum} simulations were carried out at the Leibniz Supercomputer Center (LRZ) under the project pr83li. GP acknowledges financial support from the European Research Council (ERC) under the European Union’s Horizon 2020 research and innovation program Hot- Milk (grant agreement No 865637), support from Bando per il Finanziamento della Ricerca Fondamentale 2022 dell’Istituto Nazionale di Astrofisica (INAF): GO Large program and from the Framework per l’Attrazione e il Rafforzamento delle Eccellenze (FARE) per la ricerca in Italia (R20L5S39T9). SVZ and VB acknowledge support by the \emph{Deut\-sche For\-schungs\-ge\-mein\-schaft, DFG\/} project nr. 415510302

This work is based on data from eROSITA, the soft X-ray instrument aboard SRG, a joint Russian-German science mission supported by the Russian Space Agency (Roskosmos), in the interests of the Russian Academy of Sciences represented by its Space Research Institute (IKI), and the Deutsches Zentrum für Luft- und Raumfahrt (DLR). The SRG spacecraft was built by Lavochkin Association (NPOL) and its subcontractors and is operated by NPOL with support from the Max Planck Institute for Extraterrestrial Physics (MPE).

The development and construction of the eROSITA X-ray instrument was led by MPE, with contributions from the Dr. Karl Remeis Observatory Bamberg \& ECAP (FAU Erlangen-Nuernberg), the University of Hamburg Observatory, the Leibniz Institute for Astrophysics Potsdam (AIP), and the Institute for Astronomy and Astrophysics of the University of Tübingen, with the support of DLR and the Max Planck Society. The Argelander Institute for Astronomy of the University of Bonn and the Ludwig Maximilians Universität Munich also participated in the science preparation for eROSITA.

The eROSITA data shown here were processed using the eSASS software system developed by the German eROSITA consortium.

GAMA is a joint European-Australasian project based around a spectroscopic campaign using the Anglo-Australian Telescope. The GAMA input catalogue is based on data taken from the Sloan Digital Sky Survey and the UKIRT Infrared Deep Sky Survey. Complementary imaging of the GAMA regions is being obtained by a number of independent survey programmes including GALEX MIS, VST KiDS, VISTA VIKING, WISE, Herschel-ATLAS, GMRT and ASKAP providing UV to radio coverage. GAMA is funded by the STFC (UK), the ARC (Australia), the AAO, and the participating institutions. The GAMA website is https://www.gama-survey.org/ .

%
%

\bibliographystyle{aa} 
\bibliography{laura1} 




\end{document}